# Near-infrared Emission from Defect States in Few-layer Phosphorene

*Shahriar Aghaeimeibodi [a], Je-Hyung Kim [a], and Edo Waks [a,b,]\**

[a] Department of Electrical and Computer Engineering and Institute for Research in Electronics and Applied Physics, University of Maryland, College Park, Maryland 20742, USA.

[b] Joint Quantum Institute, University of Maryland and the National Institute of Standards and Technology, College Park, Maryland 20742, USA.

\* Corresponding author E-mail: edowaks@umd.edu

ABSTRACT: Atomically-thin films of phosphorene (also known as black phosphorus) are a low-dimensional optical material with direct exciton emission, whose wavelength is tunable by controlling the number of layers. In addition to this excitonic emission, recent works revealed the existence of emissions from defect states and described new methods to manipulate them. Monolayer phosphorene exhibits emission from localized defect states at wavelengths near 920 nm. Increasing the number of layers should shift the defect emission to longer wavelengths, enabling the material to span a broader spectral range. However, defect emission from few-layer phosphorene has not yet been reported. In this paper, we demonstrate the existence of a new class of near-infrared emission from defect states in few-layer phosphorene. Photoluminescence measurements show a bright emission around 1240 nm with a sublinear growth of emission intensity when the excitation intensity is linearly increased, confirming the defect-based nature of this emission. From time-resolved lifetime measurements we determine an emission lifetime of 1.1 ns, in contrast with the exciton lifetime from phosphorene, which has been previously reported to be in the range of a few hundred picoseconds. This work shows that phosphorene defects can act as a source of infrared light with potential applications in optoelectronics.

Atomically thin phosphorene exhibits a tunable direct optical bandgap that can range from 0.35 eV (bulk) to 1.73 eV (monolayer).[1–3] This optical bandgap leads to excitonic light emission at a number of desirable infrared wavelengths that were missing from the library of optical two-dimensional materials.[2–4] These wavelengths play an important role in a broad range of optoelectronics applications, such as light emitting diodes (LEDs), optical modulators, and photo detectors.[1–3] In particular, phosphorene emission is enabling the polarization of sensitive optical devices,[5–7] which is useful for polarization dependent photo detection[8] and linearly polarized LEDs.

Besides excitonic emission, atomically thin materials also exhibit a bright emission from localized defects, which is useful for photonic and optoelectronic applications. For example, transition metal dichalcogenides (TMDs) show bright defects that even emit single photons.[9–13] More recently, several works reported strong near-infrared emission from localized defects in monolayer phosphorene at room temperature.[14,15] Multilayer phosphorene has a narrower bandgap that could shift the defect emission to the longer wavelengths required for optical communications and optoelectronic applications. However, there are no works to date reporting such longer-wavelength defect emissions, and their existence remains an open question.

In this work, we demonstrate that few-layer phosphorene possesses a new class of natural defects emitting at infrared wavelengths centered at 1240 nm. We scan the photoluminescence spectrum of three-layer phosphorene and observe localized emission from specific points on the sample surface at wavelengths that are significantly different from those of the excitons. By monitoring the intensity as a function of power, we observe a clear saturation behavior, which is indicative of defect emission. We also perform lifetime measurements that reveal a decay rate at least two times longer than that of excitons, thus verifying that the emission originates from localized defects. The defect emission is 34 times brighter than the exciton emission at low temperatures, and this brightness improvement persists up to room temperature, making defect emission adequate for room-temperature, near-infrared, optoelectronic applications including light emitting diodes, lasers, photodetectors, and optical modulators.

We prepare atomically thin flakes of phosphorene by mechanically exfoliating bulk crystals (2D Semiconductors) onto a polydimethylsiloxane thin film (Gel-Pak). We dry transfer the exfoliated flakes onto a SiO2/Si substrate with a 290 nm thick thermal oxide layer.[16] Pre-patterned alignment markers on the substrate surface serve as markers to spatially identify the location of the atomically thin flakes. To protect the flakes from the degradation caused by photo-assisted oxidation,[17] we spin-coat a 300 nm layer of polymethylmethacrylate on top of the prepared sample to encapsulate the flakes. We transfer all samples to a vacuum environment within 30 minutes of the exfoliation, to avoid prolonged exposure to the ambient environment and prevent sample degradation.

We mount the sample in a closed-cycle cryostat that reaches a minimum operating temperature of 4 K. We excite the sample and collect the emissions using a confocal microscope with an objective lens that has a numerical aperture of 0.7. The excitation is made with either continuous wave or pulsed lasers. We use a grating spectrometer with a spectral resolution of 0.3 nm to perform the spectral measurements or, alternately, a single photon counter to perform time-resolved measurements.

Figure 1a shows the optical micrograph of a region of the prepared sample surface containing few-layer phosphorene, after all the sample preparation stages. The different areas of the sample can be identified based on their color contrast. Three different regions are defined: substrate, bulk black phosphorus, and region A (which we can identify as few-layer phosphorene). We further confirm the identification of region A as three-layer phosphorene through photoluminescence measurement.

Figure 1b shows a typical photoluminescence emission spectrum from region A at 4 K when excited with a 780 nm continuous-wave laser. The photoluminescence spectrum shows a broad emission around 1062 nm, along with a sharp peak at approximately 1128 nm. By comparing this spectrum to the emission spectrum from a bare region with no phosphorene (orange curve in Figure 1b), we identify the sharp peak at 1128 nm as background Si emission. The broad emission at 1062 nm is only present in region A, and we thus attribute it to a direct emission from the phosphorene layer. We scan the beam around this region, and at a specific point in region A, we observe a drastic change of the spectrum, shown in Figure 1c, in which the emission at 1062 nm is suppressed and a bright emission at 1240 nm appears. This emission is highly localized, and quickly disappears when we move the laser focus to a slightly different location in region A.

To better understand the origin of these two distinct emission spectra, we measure the photoluminescence intensity as a function of pump power. Figure 2a shows the intensity as a function of pump power for the 1062 nm emission, calculated from the total integrated area of the photoluminescence spectrum. We subtract the narrow Si emission from the overall intensity. As shown, the emission exhibits a linear increase with pump power, which is consistent with emission from excitons. Based on the center wavelength of the excitonic emission, we infer that region A corresponds to three-layer phosphorene.[18] The solid line is the numerical fit of the function $I = a\,(P_{exc})^b + c$, where $I$ is the integrated photoluminescence count, $P_{exc}$ is the excitation power, and a, b, and c are fitting parameters. The fitting parameter $b$ determines how close the data is to linear power dependence. When this parameter is equal to unity, the intensity increases linearly with power over the entire power range, confirming that the emission requires single electron-hole pairs.[4,19] Values below unity (a sublinear increase) are indicative of a saturable emission.[14,20,21] From the fit in Figure 2a, we obtain $b = 1.04 \pm 0.1$, indicating a nearly linear behavior consistent with emission from excitons.

Figure 2b shows the integrated photoluminescence intensity as a function of pump power for the spectrum shown in Figure 1c. In contrast to the excitonic emission, this region exhibits a nonlinear intensity dependence, which is consistent with defect emission, not with emission from excitons.[14,20,21] The solid line is the numerical fit of the same function used in Figure 2a. From the fit we determine $b = 0.49 \pm 0.07$, which is sublinear and therefore consistent with emission from defects.

In transition metal dichalcogenides, the lifetimes of excitons and defects differ significantly, because of the three-dimensional quantum confinement of the localized defects.[11,13] However, the experimental demonstration of the lifetime behavior of localized defects in phosphorene is yet to be done.[22] We perform time-resolved lifetime measurements to investigate the excited state lifetime of localized defects in few-layer phosphorene. We excite the sample with a pulsed laser emitting at 780 nm with a 100 ps pulse width. We detect the emission using an avalanche photodiode (ID230 from ID Quantique) with a time resolution of 100 ps and a time interval analyzer (PicoHarp 300 from Picoquant) synchronized with the laser repetition rate. Figure 3 shows the time-resolved photoluminescence signal at 4 K, for both defect and exciton emissions.

The defect emission exhibits a longer lifetime than the exciton emission, which is consistent with the defect emission behavior of transition metal dichalcogenides.[10,13] The time-resolved photoluminescence signals for both localized defects and excitons fit well to bi-exponential curves, indicating the presence of slower processes—similar to the decay in monolayer phosphorene—in both emissions.[19] The solid curves are the numerical fits with bi-exponential decay functions. We fit both data sets to the equation $y = A_1 e^{-\frac{t}{\tau_1}} + A_2 e^{-\frac{t}{\tau_2}} + A_3$, where $y$ is the integrated counts and $t$ is the decay time. Here $A_1, A_2, A_3, \tau_1$, and $\tau_2$ are the fitting parameters. From the fit we determine a fast lifetime of $\tau_1 = 1.104$ ns and a slow lifetime of $\tau_2 = 13.9$ ns for defects and, correspondingly, $\tau_1 = 0.49$ ns and $\tau_2 = 3.7$ ns for excitons, which is in agreement with the previous measurements of excitons in monolayer phosphorene.[2,19]

Figure 4a shows the integrated intensity of the defect and excitonic emissions as a function of temperature, using an excitation power of 150 $\mu$W. At low temperatures, the defects are 34 times brighter than the excitons, and remain brighter even up to room temperature. The bright emission from the localized defects is consistent with the defect emission from monolayer phosphorene[14] and transitional metal dichalcogenides.[10–13] The brighter room-temperature emission is an important advantage for optoelectronic applications.

Figure 4b shows the full width at half maximum linewidth for both excitons and defects. As shown, the linewidths of the defect emission does not exhibit significant narrowing at low temperature, which indicates that the defect emission is inhomogeneously broadened. Figures 4c and 4d plot the center wavelength and the excited state lifetime for both the defect and excitonic emissions as a function of temperature. As shown, the center wavelength is insensitive to temperature, which is consistent with the large binding energies of excitons and defects in phosphorene.[14] The excited state lifetime of both defects and excitons are also relatively insensitive to temperature, because of the large binding energies of the system. We note here that, because of the low counts of excitonic emissions, we measure the excitonic lifetime on a different substrate, which has an extra layer of gold on top of the silicon wafer to increase the signal.

In summary, we have demonstrated a new class of infrared defect emissions from few-layer phosphorene. These defect emitters exhibit saturation and longer excited states than excitons, as is also observed in the behavior of the localized defects in TMDs. Moreover, they are 34 times brighter than the excitons at low temperature, and persist to be brighter even at room temperature. They could be used as room temperature sources of near-infrared emission for optoelectronics, especially for applications in the fields of light emitting diodes, lasers, and polarization-sensitive detection. We observed significant inhomogeneous broadening in these defect emitters. Tunable and robust near-infrared emitters can possibly be obtained by introducing the defects intentionally with any of the reported techniques for defect engineering in phosphorene and other low-dimensional materials, such as substrate modification, oxygen plasma etching,[14,15] strain, and electron-beam irradiation.[23]

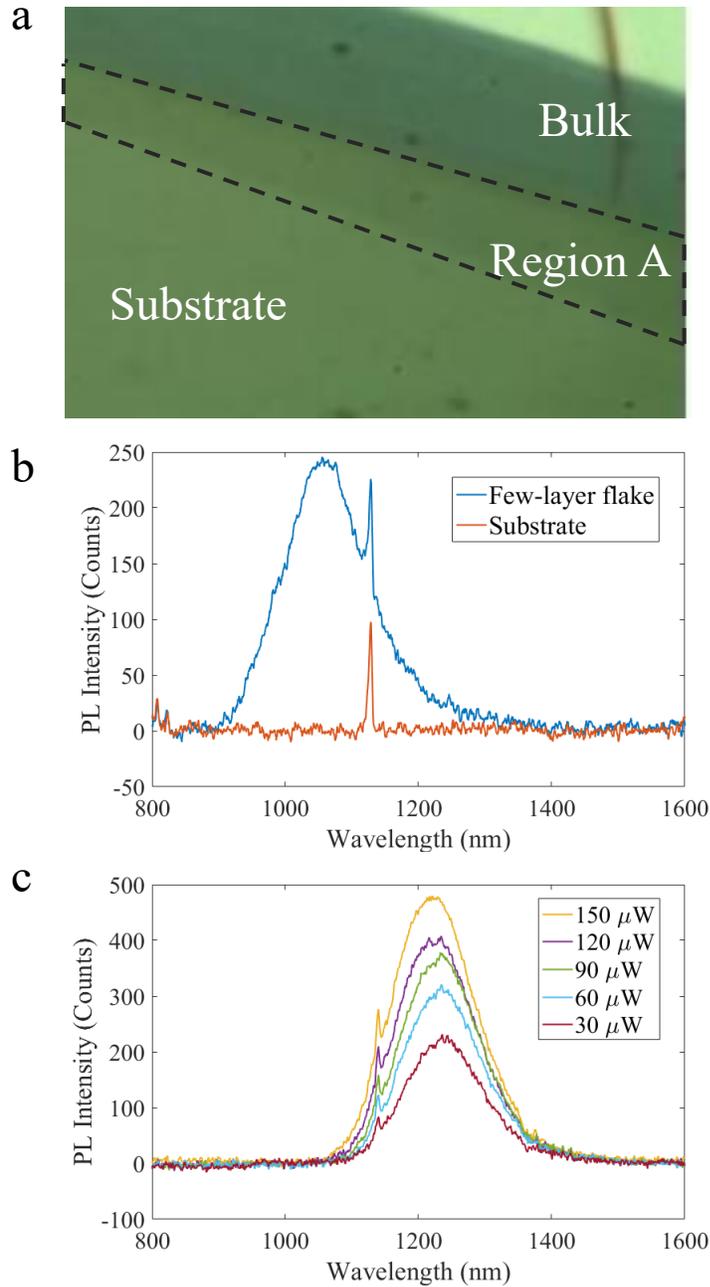

Figure 1. (a) Optical micrograph of the exfoliated few-layer phosphorene after all the sample preparation steps. The substrate, few-layer phosphorene (region A), and bulk regions are marked in the graph. (b) Typical photoluminescence of region A (in blue) and substrate (in orange), at 4 K. (c) Photoluminescence of a localized emission located in region A, at 4 K.

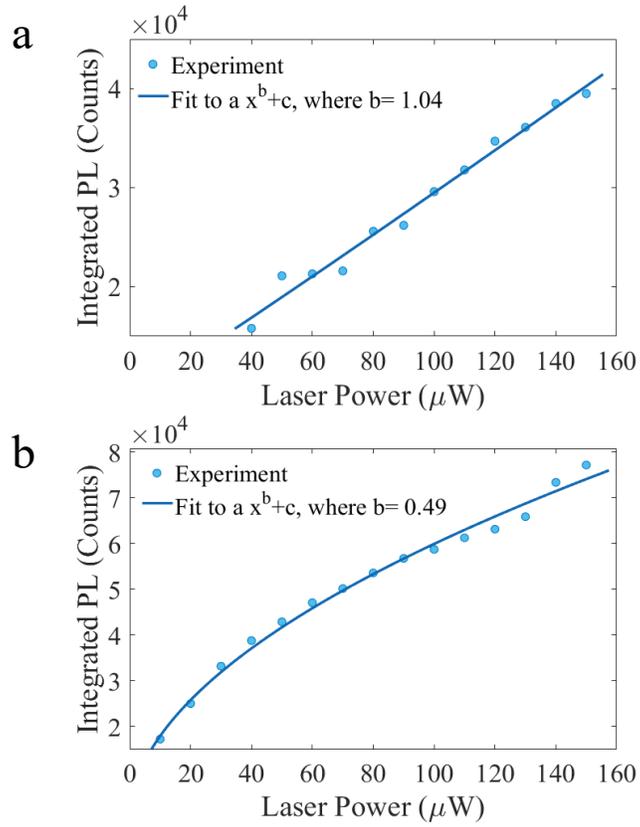

Figure 2. Power dependence of the integrated photoluminescence emission after elimination of the substrate emission. The solid points are the measured data, and the blue curve is the power law fitting curve. (a) Photoluminescence emission in Figure 1b. (b) Photoluminescence emission in Figure 1c.

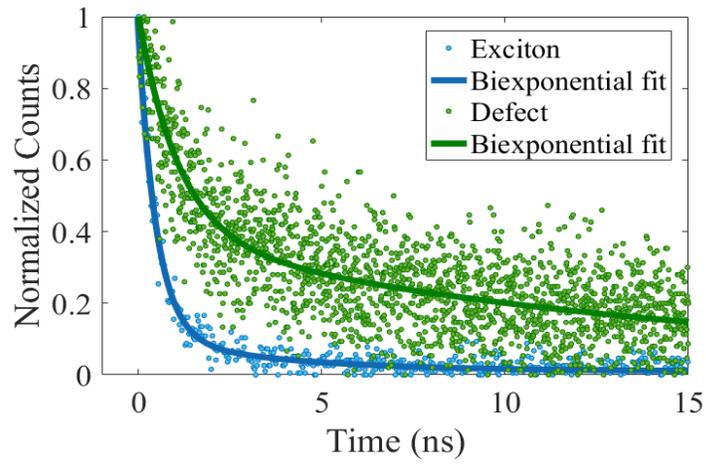

Figure 3. Time-resolved photoluminescence of the few-layer flake at 4 K for localized defects (Green) and excitons (Blue). The solid curves are the bi-exponential fits to the data.

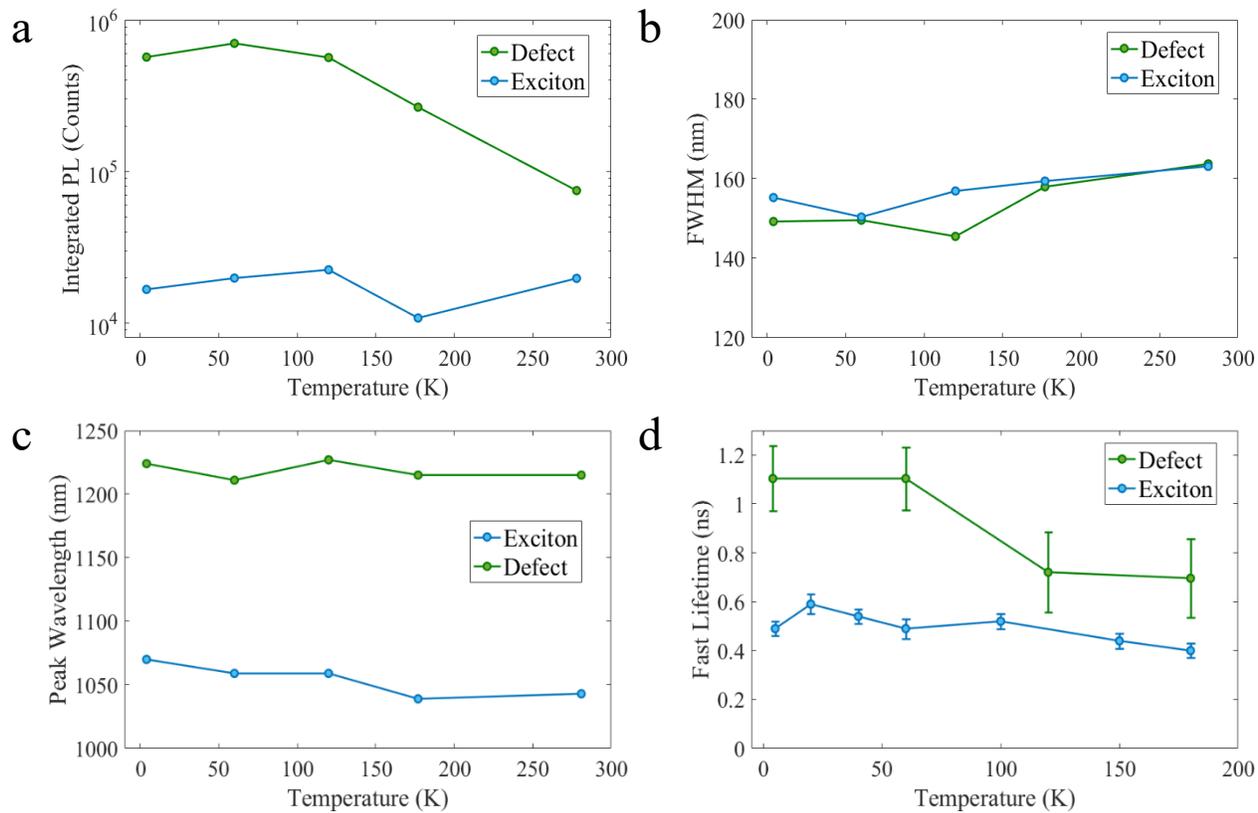

Figure 4. Temperature dependence of the photoluminescence for defects (green) and excitons (blue). (a) Integrated photoluminescence intensity. (b) Full width at half maximum. (c) Peak energy. (d) Fast lifetime of exciton and defects measured with time-resolved photoluminescence measurements.

AUTHOR INFORMATION

**Corresponding Author**

*E-mail: edowaks@umd.edu.

**Funding Sources**

The authors acknowledge support from an Office of Naval Research ONR grant (award number N000141410612), and an AFOSR Center of Excellence for Science of Electronics in Extreme Electromagnetic Environments (award number 271470871D).

**Notes**

The authors declare no competing financial interest.

REFERENCES

(1) Li, L.; Kim, J.; Jin, C.; Ye, G. J.; Qiu, D. Y.; da Jornada, F. H.; Shi, Z.; Chen, L.; Zhang, Z.; Yang, F.; Watanabe, K.; Taniguchi, T.; Ren, W.; Louie, S. G.; Chen, X. H.; Zhang, Y.; Wang, F. *Nat. Nanotechnol.* **2017**, *12* (1), 21–25.

(2) Yang, J.; Xu, R.; Pei, J.; Myint, Y. W.; Wang, F.; Wang, Z.; Zhang, S.; Yu, Z.; Lu, Y. *Light Sci. Appl.* **2015**, *4* (7), e312.

(3) Tran, V.; Soklaski, R.; Liang, Y.; Yang, L. *Phys. Rev. B - Condens. Matter Mater. Phys.* **2014**, *89* (23), 235319.

(4) Surrente, A.; Mitioglu, A. A.; Galkowski, K.; Tabis, W.; Maude, D. K.; Plochocka, P. *Phys. Rev. B - Condens. Matter Mater. Phys.* **2016**, *93* (12), 121405.

(5) Wang, X.; Jones, A. M.; Seyler, K. L.; Tran, V.; Jia, Y.; Zhao, H.; Wang, H.; Yang, L.; Xu, X.; Xia, F. *Nat. Nanotechnol.* **2015**, *10* (6), 517–521.

(6) Qiao, J.; Kong, X.; Hu, Z.-X.; Yang, F.; Ji, W. *Nat. Commun.* **2014**, *5*, 4475.

(7) Xia, F.; Wang, H.; Jia, Y. *Nat. Commun.* **2014**, *5*, 4458.

(8) Yuan, H.; Liu, X.; Afshinmanesh, F.; Li, W.; Xu, G.; Sun, J.; Lian, B.; Curto, A. G.; Ye, G.; Hikita, Y.; Shen, Z.; Zhang, S.-C.; Chen, X.; Brongersma, M.; Hwang, H. Y.; Cui, Y. *Nat. Nanotechnol.* **2015**, *10* (8), 703–713.

(9) Tran, T. T.; Bray, K.; Ford, M. J.; Toth, M.; Aharonovich, I. *Nat. Nanotechnol.* **2015**, 11, 37–41

(10) Koperski, M.; Nogajewski, K.; Arora, A.; Cherkez, V.; Mallet, P.; Veuillen, J.-Y.; Marcus, J.; Kossacki, P.; Potemski, M. *Nat. Nanotechnol.* **2015**, *10* (6), 503–506.

(11) He, Y.-M.; Clark, G.; Schaibley, J. R.; He, Y.; Chen, M.-C.; Wei, Y.-J.; Ding, X.; Zhang, Q.; Yao, W.; Xu, X.; Lu, C.-Y.; Pan, J.-W. *Nat. Nanotechnol.* **2015**, *10* (6), 497–502.

(12) Chakraborty, C.; Kinnischtzke, L.; Goodfellow, K. M.; Beams, R.; Vamivakas, A. N. *Nat. Nanotechnol.* **2015**, *10* (6), 507–511.

(13) Srivastava, A.; Sidler, M.; Allain, A. V.; Lembke, D. S.; Kis, A.; Imamoğlu, A. *Nat. Nanotechnol.* **2015**, *10* (6), 491–496.

(14) Xu, R.; Yang, J.; Myint, Y. W.; Pei, J.; Yan, H.; Wang, F.; Lu, Y. *Adv. Mater.* **2016**, *28* (18), 3493–3498.

(15) Pei, J.; Gai, X.; Yang, J.; Wang, X.; Yu, Z.; Choi, D.-Y.; Luther-Davies, B.; Lu, Y. *Nat. Commun.* **2016**, *7*, 10450.

(16) Castellanos-Gomez, A.; Buscema, M.; Molenaar, R.; Singh, V.; Janssen, L.; van der Zant, H. S. J.; Steele, G. A. *2D Mater.* **2014**, *1* (1), 11002.

(17) Favron, A.; Gaufrès, E.; Fossard, F.; Phaneuf-L'Heureux, A.-L.; Tang, N. Y.-W.; Lévesque, P. L.; Loiseau, A.; Leonelli, R.; Francoeur, S.; Martel, R. *Nat. Mater.* **2015**, *14* (8), 826–832.

(18) Xu, R.; Zhang, S.; Wang, F.; Yang, J.; Wang, Z.; Pei, J.; Myint, Y. W.; Xing, B.; Yu, Z.; Fu, L.; Qin, Q.; Lu, Y. *ACS Nano* **2016**, *10* (2), 2046–2053.

(19) Surrente, A.; Mitioglu, A. A.; Galkowski, K.; Klopotowski, L.; Tabis, W.; Vignolle, B.; Maude, D. K.; Plochocka, P. *Phys. Rev. B - Condens. Matter Mater. Phys.* **2016**, *94* (7), 075425.

(20) You, Y.; Zhang, X.-X.; Berkelbach, T. C.; Hybertsen, M. S.; Reichman, D. R.; Heinz, T. F. *Nat. Phys.* **2015**, *11* (6), 477–481.

(21) Shang, J.; Shen, X.; Cong, C.; Peimyoo, N.; Cao, B.; Eginligil, M.; Yu, T. *ACS Nano* **2015**, *9* (1), 647–655.

(22) Long, R.; Fang, W.; Akimov, A. V. *J. Phys. Chem. Lett.* **2016**, *7* (4), 653–659.

(23) Tran, T. T.; Elbadawi, C.; Totonjian, D.; Lobo, C. J.; Grosso, G.; Moon, H.; Englund, D. R.; Ford, M. J.; Aharonovich, I.; Toth, M. *ACS Nano* **2016**, *10* (8), 7331–7338.

Table of Contents Graphic

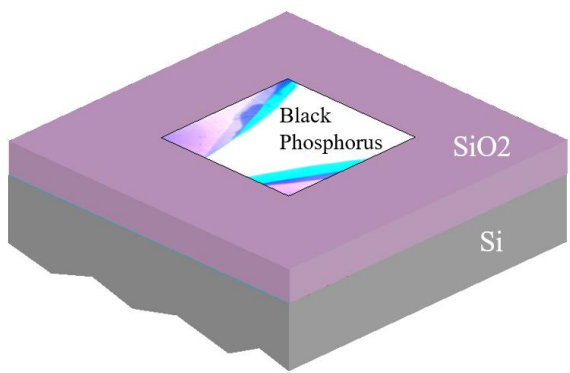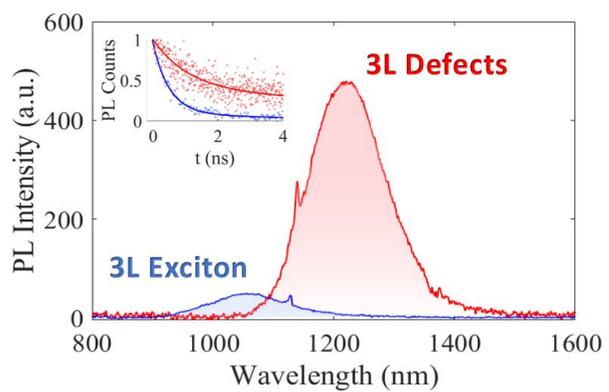